\documentclass[reprint, superscriptaddress, longbibliography, floatfix, twocolumn, nofootinbib, amsmath, amssymb, aps, showkeys]{revtex4-2}
\usepackage{mathtools}
\usepackage{nicefrac}
\usepackage{braket}
\usepackage{overpic}
\usepackage{siunitx}
\usepackage{hyperref}
\usepackage{gensymb}
\usepackage{xcolor}
\usepackage{graphicx}
\usepackage{dcolumn}
\usepackage{bm}
\usepackage{dsfont}

\newcommand{\appref}[1]{\hyperref[#1]{Appendix~\ref*{#1}}}
\makeatletter
\newcommand*{\rom}[1]{\expandafter\@slowromancap\romannumeral #1@}
\makeatother

\begin{document}
\preprint{APS/123-QED}

\title{
Telecom-compatible polarization-to-time-bin conversion of \\ 
atom-photon entanglement for heterogeneous quantum networks}


\author{Christian Haen}
\affiliation{%
Fachrichtung Physik, Universit\"at des Saarlandes, 66123 Saarbr\"ucken, Germany 
}%
\affiliation{%
Zentrum für Quantentechnologien (QuTe), Universität des Saarlandes, 66123 Saarbr\"ucken, Germany 
}%

\author{Julian Gro\ss-Funk}
\thanks{Current address: Institut für Quantentechnologien, DLR, 89081 Ulm, Germany}
\affiliation{%
Fachrichtung Physik, Universit\"at des Saarlandes, 66123 Saarbr\"ucken, Germany 
}%

\author{Max Bergerhoff}
\affiliation{%
Fachrichtung Physik, Universit\"at des Saarlandes, 66123 Saarbr\"ucken, Germany 
}%
\affiliation{%
Zentrum für Quantentechnologien (QuTe), Universität des Saarlandes, 66123 Saarbr\"ucken, Germany 
}%

\author{Pascal Baumgart}
\affiliation{%
Fachrichtung Physik, Universit\"at des Saarlandes, 66123 Saarbr\"ucken, Germany 
}%
\affiliation{%
Zentrum für Quantentechnologien (QuTe), Universität des Saarlandes, 66123 Saarbr\"ucken, Germany 
}%

\author{Jonas Meiers}
\affiliation{%
Fachrichtung Physik, Universit\"at des Saarlandes, 66123 Saarbr\"ucken, Germany 
}%
\affiliation{%
Zentrum für Quantentechnologien (QuTe), Universität des Saarlandes, 66123 Saarbr\"ucken, Germany 
}%

\author{Tobias Bauer}
\affiliation{%
Fachrichtung Physik, Universit\"at des Saarlandes, 66123 Saarbr\"ucken, Germany 
}%
\affiliation{%
Zentrum für Quantentechnologien (QuTe), Universität des Saarlandes, 66123 Saarbr\"ucken, Germany 
}%

\author{Christoph Becher}
\affiliation{%
Fachrichtung Physik, Universit\"at des Saarlandes, 66123 Saarbr\"ucken, Germany 
}%
\affiliation{%
Zentrum für Quantentechnologien (QuTe), Universität des Saarlandes, 66123 Saarbr\"ucken, Germany 
}%

\author{J\"urgen Eschner}
\email{juergen.eschner@physik.uni-saarland.de}
\affiliation{%
Fachrichtung Physik, Universit\"at des Saarlandes, 66123 Saarbr\"ucken, Germany 
}%
\affiliation{%
Zentrum für Quantentechnologien (QuTe), Universität des Saarlandes, 66123 Saarbr\"ucken, Germany 
}%

\date{\today}

\begin{abstract}

A key enabling feature of future quantum networks is interoperability  between platforms that operate at different wavelengths and with different qubit encodings. We demonstrate an interface that converts atom-photon entanglement from polarization encoding at an atomic wavelength to time-bin encoding in the telecom C-band. Atom-entangled photons at 854\,nm are generated from a single $^{40}$Ca$^+$ ion. After quantum frequency conversion to 1550\,nm, the photonic polarization qubit is converted into a time-bin qubit using a fiber-based Mach–Zehnder-like encoder. Full quantum tomography of the final state verifies that the process preserves entanglement with 96.3(4.2)\% fidelity.
Together with the independent work of Ferrari et al.\ [arXiv:2607.07805 (2026)], this is the first demonstration of polarization-to-time-bin conversion of photons entangled with a single atomic quantum memory. The telecom-compatible interface enables robust qubit transmission over optical fibers and provides a key building block for heterogeneous quantum networking architectures.
\end{abstract}


\maketitle
\section{Introduction}
For the realization of large-scale quantum networks \cite{Kimble_2008} and distributed quantum computing \cite{DiVincenzo_2000, Nigmatullin_2016, Jiang_2007, Cirac_1999}, quantum information must be distributed between distant quantum memories, or processors, using photonic qubits. 
In addition, future quantum networks are expected to combine different physical platforms, each optimized for specific tasks such as long-lived storage, efficient spin-photon interfaces, or scalable integration \cite{vanLoock_2020, Wei_2022}. Consequently, their photonic interfaces must bridge differences in wavelength, bandwidth, wave-packet shape, and also in qubit encoding. While trapped ions and neutral atoms naturally establish memory--photon entanglement in the polarization degree of freedom \cite{Bock_2018, vanLeent_2022}, several solid-state platforms employ time-bin-encoded photonic qubits \cite{Jayakumar_2014, Tchebotareva_2019, Knaut_2024}. Coherent conversion between polarization and time-bin encoding is therefore an important enabling technology for heterogeneous quantum networks.

Telecom optical fibers provide readily available infrastructure for metropolitan- and long-distance quantum communication, where transmission losses may ultimately be overcome using quantum repeaters \cite{Briegel_1998}. Polarization-encoded qubits, however, are susceptible to environmental perturbations during fiber transmission. Temperature variations and mechanical stress modify the fiber birefringence and result in time-dependent polarization transformations \cite{Ulrich_1979, Ding_2017, Kucera_2024}. Deployed fiber links therefore generally require active polarization stabilization or compensation \cite{Kucera_2024}. In time-bin encoding, the early and late components traverse the fiber within a short temporal separation and experience approximately the same polarization transformation, making the encoded information insensitive to sufficiently slow polarization fluctuations. Consequently, while quantum communication has been successfully demonstrated with photonic polarization qubits, time-bin qubits may be considered the more robust and more widely applicable solution. Moreover, in contrast to direct generation of time-bin qubits from quantum memories \cite{Ward_2022, Saha_2025}, polarization-to-time-bin conversion allows the temporal separation of the time bins to be determined by an external interferometer rather than by the emitter dynamics. This flexibility facilitates the interfacing of quantum memories with different emission timescales and control sequences. Hence, conversion of polarization to time-bin qubits may become a standard element in quantum networks, in addition to quantum frequency conversion. 

Conversion of polarization to time-bin qubits has previously been implemented using both active \cite{Kupchak_2017} and passive schemes \cite{Sanaka_2002}. Active approaches can avoid the intrinsic loss associated with passive recombination, but require fast optical switching synchronized with the photon arrival time, limiting their application to single photon sources with controlled emission times. Passive conversion operates independently of the photon arrival time, at the expense of an intrinsic efficiency penalty. Passive conversion schemes have been employed with attenuated laser pulses in BB84 quantum key distribution \cite{Scalcon_2022}, with polarization-entangled photon pairs generated by spontaneous parametric down-conversion \cite{Martin_2013}, as well as with photons emitted by quantum dots, where polarization-to-time-bin conversion was integrated with telecom quantum frequency conversion \cite{Yu_2015}. The reverse conversion from time-bin to polarization encoding has also been demonstrated \cite{Sanaka_2002, Takesue_05, Bussieres_2010}, including for a photon entangled with a spin in a nitrogen vacancy color center \cite{Vasconcelos_2020}. Those experiments established coherent photonic encoding conversion, but did not demonstrate polarization-to-time-bin conversion for a photon entangled with an atomic quantum memory.

In this work, we demonstrate a telecom C-band-compatible interface for a $^{40}$Ca$^+$ trapped-ion quantum node. 
Single photons emitted by the ion and entangled with it \cite{Bock_2018} are converted from polarization qubits at the ionic transition wavelength ($854\,$nm) to time-bin qubits in the telecom C-band ($1550\,$nm), using quantum frequency conversion (QFC) \cite{Arenskoetter_2023} and subsequent polarization-to-time-bin conversion. The final entanglement is verified by quantum tomography.  

In a parallel and independent study, Ferrari et al.\ demonstrate 
polarization-to-time-bin conversion of photons entangled with a $^{88}\mathrm{Sr}^{+}$ ion \cite{Ferrari_2026}. 
While Ferrari et al. show that atom-photon entanglement is preserved in the conversion and characterize the robustness of the converted state against externally applied polarization noise using fidelity bounds from partial basis measurements, our work establishes compatibility with long-distance quantum networks through bidirectional quantum frequency conversion and reconstructs the complete density matrix and entanglement fidelity of the converted state via full quantum state tomography. 
We discuss the potential application of the demonstrated qubit interface to the Saarbrücken urban fiber network \cite{Kucera_2024} as a key enabling technology for a heterogeneous quantum link, combining trapped ions with color centers in diamond \cite{Schaefer_2025, Herrmann_2026}.

\section{Experiment}\label{sec:Experiment}

\subsection{Experimental setup}

\begin{figure*}[t]
		\centering
		\begin{overpic}[width=\linewidth]{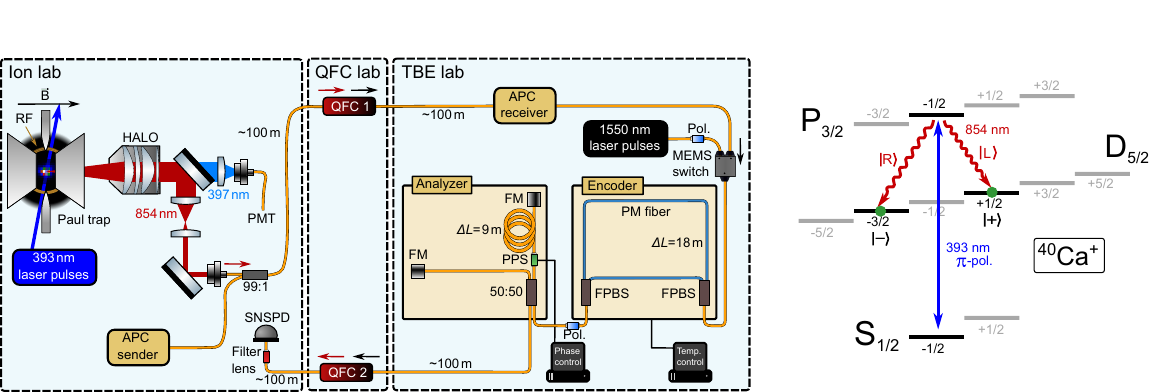}
			\put(0,30.5){\Large a)}
            \put(67,30.5){\Large b)}
		\end{overpic}
\caption{
(a) Schematic of the interface converting 854-nm polarization-encoded photons to 1550-nm time-bin-encoded photons including the setup for generation and analysis of atom-photon entanglement. (b) Relevant level scheme of $^{40}$Ca$^+$ and atom-photon entanglement generation. Detailed descriptions are given in the text. RF: radio frequency; HALO: high-numerical-aperture laser objective; PMT: photomultiplier tube; APC: automated polarization drift compensation; SNSPD: superconducting nanowire single-photon detector; QFC: quantum frequency conversion; TBE: time-bin encoder; MEMS: microelectromechanical system; PM: polarization maintaining; FPBS: fiber polarizing beam splitter; FM: Faraday mirror; PPS: piezo phase shifter.}
\label{fig:setup}
\end{figure*}

In the experiment, shown schematically in \autoref{fig:setup}\,a), a single $^{40}$Ca$^+$ ion is trapped in a linear Paul trap; relevant atomic levels and transitions are shown in \autoref{fig:setup}\,b). Lasers at $397\,$nm and $866\,$nm (not shown) are used for Doppler cooling and state discrimination via fluorescence detection, while a 729-nm laser enables coherent excitation and electron shelving to the D$_{5/2}$ metastable state. A 393-nm laser beam excites the ion to create single 854-nm photons \cite{Baumgart_2026}. The 393-nm light is shaped into $\sim 50$-ns pulses by an acousto-optic modulator (AOM), thus setting the time scale for the photon wave-packet. Emitted 854-nm photons are collected with an in-vacuum high-numerical-aperture laser objective (HALO) and coupled into a single-mode fiber. Photons are collected along the quantization axis, which is defined by a static magnetic field of $\sim 2.85\,$G. Fluorescence photons at $397\,$nm are coupled into a multi-mode fiber and detected using a photomultiplier tube (PMT). More details of the trap setup are found in \cite{Bergerhoff_2024}.

The fiber-coupled 854-nm photons are sent through $\sim100\,$m of fiber to a polarization-maintaining quantum frequency converter (QFC) \cite{Arenskoetter_2023} in a separate lab. The resulting $1550$-nm photons are guided through an additional $\sim100\,$m of fiber to a third lab into the time-bin encoder (TBE) and subsequently to an analyzer; both are described in detail in \autoref{sec:TBE}. Since the fibers connecting the labs run through non–temperature-stabilized hallways, an automated polarization drift compensation (APC) system is implemented between the trap and the TBE. The APC is a modified version of the setup introduced in \cite{Kucera_2024}, here using a polarization reference at $854\,$nm. Slow polarization drifts in fibers not compensated by the APC are corrected periodically as described in \appref{app:slow_pol_comp}. After passing through the analyzer, the photons are transmitted to a second QFC \cite{Bauer_2023} for their detection at $854\,$nm with a superconducting nanowire single-photon detector (SNSPD). A free-space filter lens reduces conversion noise before detection.

To establish a reference for the quantum state fidelity without time-bin conversion, the TBE and analyzer are bypassed, and the back-converted photons are analyzed using a polarization projection setup as described in \cite{Bergerhoff_2024}.

\subsection{Atom-photon entanglement}

Atom–photon entanglement is generated according to the protocols in \cite{Bock_2018, Bock_2024, Bergerhoff_2024, Bergerhoff_2026}. The relevant energy levels are shown in \autoref{fig:setup}\,b). The ion is initialized in the ground state S$_{1/2}$ and subsequently excited by a $\pi$-polarized 393-nm laser pulse, leading to the emission of a photon at $854\,$nm.
Only the decay channels populating the Zeeman sub-levels $\ket{+}=\ket{D_{5/2},+1/2}$ and $\ket{-}=\ket{D_{5/2},-3/2}$ are used for the generation of entanglement. These two channels are selected through the atomic state projection performed at the end of the protocol, allowing us to filter out all other decay paths. The emitted photon is therefore polarization-entangled with the ionic Zeeman state. The difference in the Clebsch–Gordan coefficients is balanced by introducing a controlled population loss, resulting (ideally) in the maximally entangled atom–photon state
\begin{align}
    \ket{\psi}  &= \sqrt{\frac{1}{2}} \left( \ket{+} \ket{L} + e^{i \omega_L t } \ket{-} \ket{R} \right), 
    \label{equ:ape} 
\end{align}
where $\ket{L}$ and $\ket{R}$ are left- and right-hand circularly polarized single-photon states, $\omega_L = 2\pi \times 9.6$\,MHz represents the Larmor frequency between $\ket{+}$ and $\ket{-}$, and $t$ the time elapsed after emission of the photon. 

The 854-nm photon is transformed to 1550\,nm by quantum frequency conversion, maintaining the photonic polarization and thereby the entanglement of \autoref{equ:ape} \cite{Bock_2024, Bergerhoff_2026}. 

\subsection{Time-bin conversion and analysis}\label{sec:TBE}

The time-bin encoder is an unbalanced Mach-Zehnder fiber interferometer with a path length difference of $\Delta L\approx 18\,$m, corresponding to a travel time difference of $\sim 90\,$ns. It consists of two fiber polarization beam splitters (FPBS) connected by polarization-maintaining SMF-28 telecom fibers. The horizontal polarization component ($\ket{H}$) is transmitted along the short path, while the vertical component ($\ket{V}$) is sent through the long fiber. A polarizer is placed behind the interferometer and set to diagonal linear polarization ($\ket{D}$), in order to erase which-way information. Thus the encoder converts a polarization input state
\begin{align}
    \ket{\Psi_1}=\frac{1}{\sqrt{N}}\left(\alpha\ket{H}+\beta \ket{V}\right),  
\end{align}
with amplitudes $\alpha$ and $\beta$ and normalization constant $N$, into the time-bin output state 
\begin{align}\label{equ:encoder}
    \ket{\Psi_2}=\frac{1}{\sqrt{N}}\left(\alpha\ket{1}+\beta e^{i\phi_1}\ket{2}\right).  
\end{align}
Here, $\ket{1}$ and $\ket{2}$ denote the 
early and late time-bin states, and $\phi_1$ is the additional phase introduced by the travel time difference between both paths. 

\begin{figure*}[t]
		\centering
		\begin{overpic}[width=0.49\linewidth]{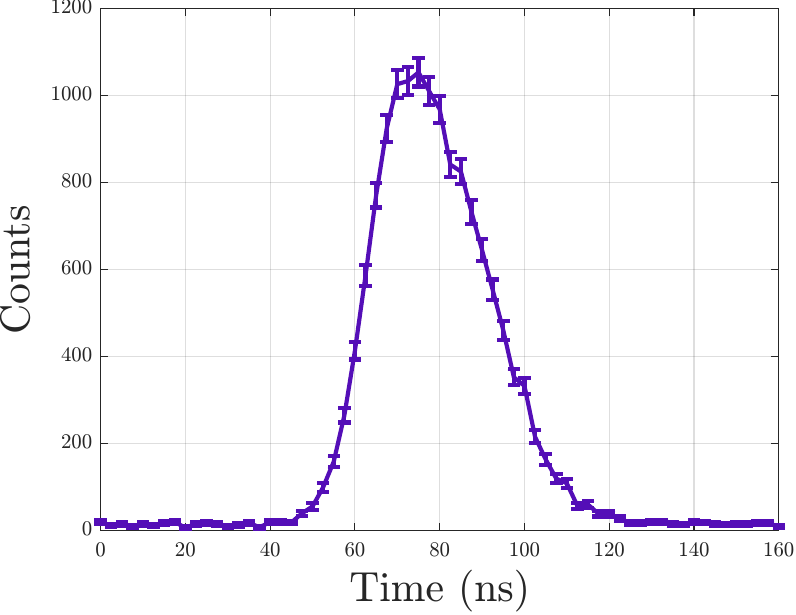}
			\put(0,80){\Large a)}
		\end{overpic}%
		\begin{overpic}[width=0.49\linewidth]{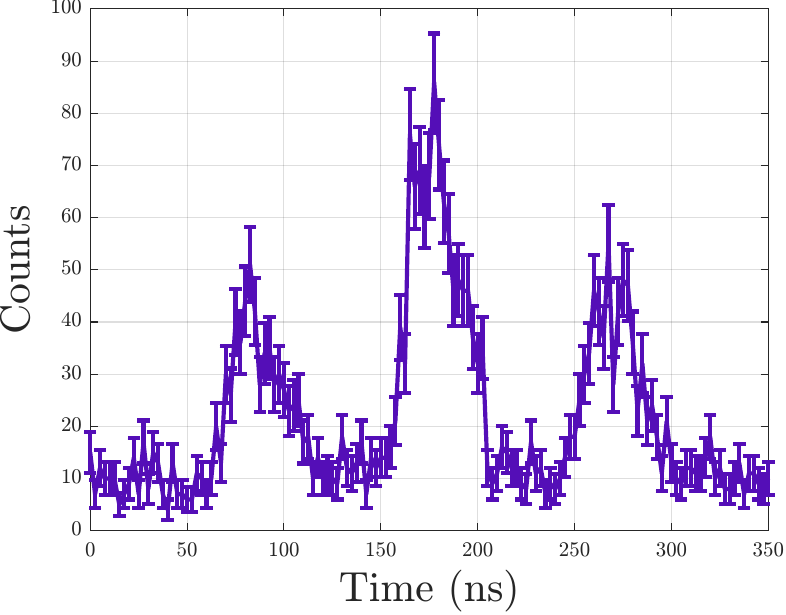}
			\put(0,80){\Large b)}
		\end{overpic}
		\caption{Measured photon wave-packets with a bin size of $2.5\,$ns, a) without time-bin encoding (integration time $4.5\,$h) b) with time-bin encoding and subsequent analysis (integration time $10\,$h).}
\label{fig:Photons}
\end{figure*}

The interferometer is enclosed in a thermally insulated housing for passive temperature stability. In addition, the temperature is monitored and actively stabilized using a PID controller driving a resistive heating wire. The temperature sensor, heating wire, and TBE fibers are mounted on a common aluminum plate to minimize temperature gradients. Before a measurement, the stabilization system is operated for several hours to ensure thermal equilibrium. This results in a long-term temperature stability better than $1\,\mathrm{mK}$, as characterized in \appref{app:temp_stab}.

After conversion to time-bins, the photon state is analyzed by a second unbalanced interferometer, of Michelson-type. The incoming signal is divided by a 50:50 non-polarizing beam splitter into a short and long fiber path, whose length difference matches the previous time difference of $90\,$ns. Both fibers are terminated by Faraday mirrors that reflect the light back to the input beam splitter. The beam splitter then recombines both paths into an output fiber. This set-up reduces the influence of birefringence and polarization-dependent loss \cite{Muller_1997}. It results in three possible arrival-time states, $\ket{1}$, $\ket{2}$, and $\ket{3}$, with a respective delay of 0\,ns, 90\,ns, and 180\,ns. The states $\ket{1}$ and $\ket{3}$ occur when the previous early or late time-bin photons travel again, respectively, through the short or long analyzer path. The state $\ket{2}$ corresponds to two amplitudes, either the early time-bin entering the long analyzer path, or vice versa. This causes interference at the beam splitter. The resulting state is given by
\begin{align}
\label{equ:analyzer}
    \ket{\Psi_3}=\frac{1}{\sqrt{N^{\prime}}}\left(\alpha\ket{1}+\left(\alpha  e^{i\phi_2}+\beta e^{i\phi_1}\right)\ket{2}+\beta e^{i(\phi_1+\phi_2)}\ket{3}\right),  
\end{align}
with phase $\phi_2$ caused by traversing the long analyzer path, and normalization constant $N^{\prime}$. We control $\phi_2$ via a piezo-based phase shifter in the long path. 
Measurement of a photon in state $\ket{\Psi_3}$ leads to a detection event in one of the three arrival-time bins. The photon is projected onto horizontal or vertical polarization if it is found in $\ket{1}$ or $\ket{3}$, respectively. A photon found in $\ket{2}$ is projected onto one of the other basis polarizations $\ket{D}$, $\ket{A}$, $\ket{R}$, or $\ket{L}$, depending on the set phase $\phi_2$, which controls the interference between the two amplitudes contributing to $\ket{2}$. 

To reduce thermally induced phase fluctuations, the analyzer employs both passive and active phase stabilization. Passive stability is gained by enclosing the analyzer in a thermally insulated housing, similar to that used for the TBE, and by replacing a section of the long interferometer arm with a fiber exhibiting a negative thermal coefficient of delay of $-10\, \mathrm{ps\, km^{-1}\,K^{-1}}$ (Linden-SPE-7192). Active phase stabilization (of $\phi_2-\phi_1$) is implemented using the piezo-based phase shifter together with an attenuated reference laser injected via a MEMS fiber switch before the TBE interferometer. This is described in detail in \appref{app:phase_stab}. Characterization of the losses of the TBE and analyzer is provided in \appref{app:losses}. 

\section{Results}

\begin{figure*}[t]
		\centering
		\begin{overpic}[width=\linewidth]{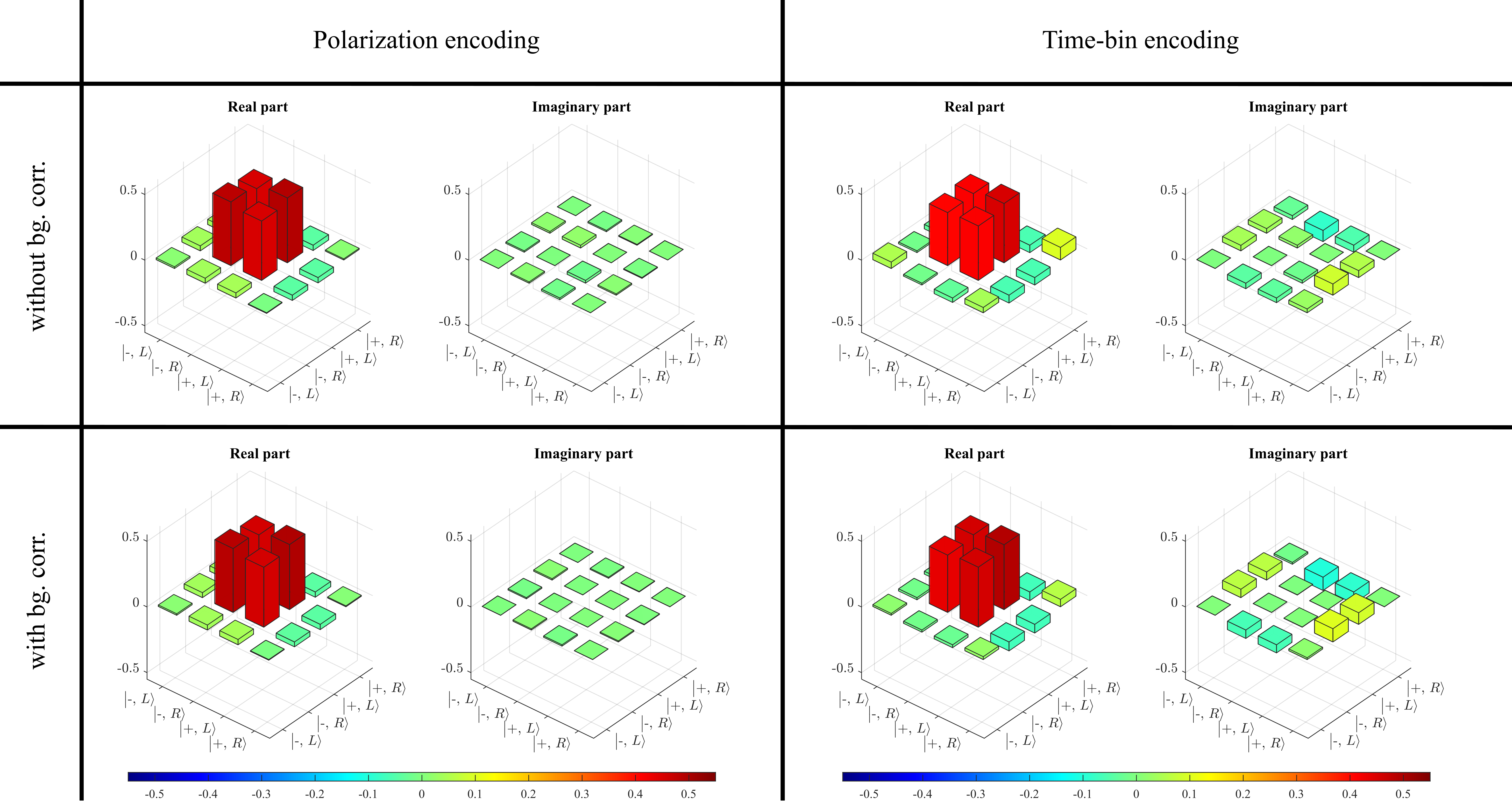}
			\put(7,44.5){\Large a)}
            \put(53.3,44.5){\Large b)}
            \put(7,21.5){\Large c)}
            \put(53.3,21.5){\Large d)}
		\end{overpic}
		    \caption{Quantum-state tomography: real and imaginary part of the density matrices of the ion–photon entangled state with double quantum frequency conversion, a) and c) with polarization encoding, without and with background correction, b) and d) with time-bin encoding and subsequent analyzer, without and with background correction. The fidelities of the states in panels a) - d) to the maximally entangled state are $94.1(0.8)\%$, $83.3(4.8)\%$, $94.8(0.8)\%$, and $91.3(4.1)\%$, respectively.}
\label{fig:DensityMatrices}
\end{figure*}
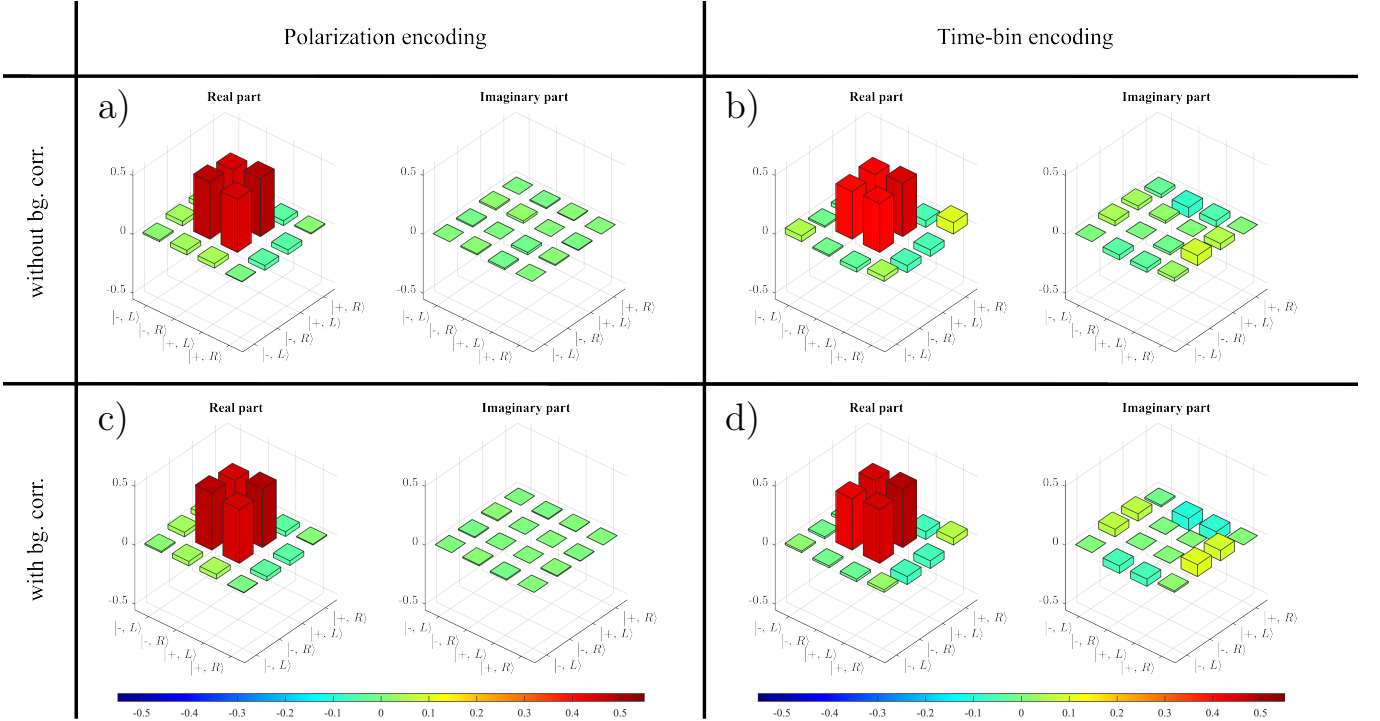

In order to assess the fidelity of our polarization-to-time-bin converter, we first evaluate the fidelities of individual parts of the setup. 

We determine the contribution of the combined TBE and analyzer interferometers to the overall atom-photon entanglement fidelity by measuring their process fidelity $\mathcal{F}_\mathrm{P}$. Using attenuated 1550-nm laser pulses, we obtain $\mathcal{F}_\mathrm{P}=97.3(1.1)\%$, as described in \appref{app:conv_fid}. For a photon of the maximally entangled state of \autoref{equ:ape} that is transmitted through TBE and analyzer, the remaining entanglement fidelity is equal to $\mathcal{F}_\mathrm{P}$ \cite{schumacher1996,Vardoyan_2022}. 

In a next step, we establish a reference value for the fidelity of the atom-photon polarization entanglement, $\mathcal{F}_\mathrm{pol}$, after quantum frequency conversion to $1550\,$nm and back to $854\,$nm, but without time-bin conversion. In order to maintain the polarization fidelity between atom and detector above 99\%, APC is activated every $80\,$s with the APC receiver placed directly before the second QFC. $\mathcal{F}_\mathrm{pol}$ is then evaluated by quantum state tomography using maximum likelihood estimation \cite{James_2001, Bock_2018}. The reconstructed density matrix, shown in \autoref{fig:DensityMatrices}$\,$a), yields a raw fidelity of $\mathcal{F}_\mathrm{pol,bg}=94.1(0.8)\%$ to the state of \autoref{equ:ape}. The corresponding photon wave-packet is shown in \autoref{fig:Photons}\,a), with a signal to background ratio (SBR) of $54.7$. Applying the background-correction procedure introduced in \cite{Bock_2018} results in a background-corrected fidelity of $\mathcal{F}_\mathrm{pol}=94.8(0.8)\%$, with the corresponding density matrix depicted in \autoref{fig:DensityMatrices}~c).


We then insert the TBE and analyzer interferometers after the first QFC, remove the polarization projection setup, and place the APC receiver in front of the TBE. The APC is activated every $80\,$s, and the analyzer phase (i.e., $\phi_2-\phi_1$) is actively stabilized every $10\,$s. We again perform quantum state tomography, this time on the time-bin detections, resulting in a fidelity of $\mathcal{F}_\mathrm{TB,bg}=83.3(4.8)\%$ with background. The average SBR over all three time bins amounts to $3.5$, as illustrated in \autoref{fig:Photons}\,b). Applying background correction yields a fidelity of $\mathcal{F}_\mathrm{TB}=91.3(4.1)\%$. The fidelity reduction from polarization encoding at 854\,nm to time-bin encoding at 1550\,nm by $\mathcal{F}_\mathrm{TB} / \mathcal{F}_\mathrm{pol} = 96.3(4.2)\%$ is in good agreement with the measured process fidelity of TBE and analyzer, $\mathcal{F}_\mathrm{P}$. The density matrices of the time-bin-encoded atom-photon states without and with background correction are shown in \autoref{fig:DensityMatrices}$\,$b) and d). The SBR-limited fidelity is mainly a result of the significant losses in the TBE and analyzer.

\section{Summary and Discussion}

We have realized and characterized an interface that converts atom-photon polarization entanglement created at the atomic wavelength of 854\,nm to time-bin entanglement at the telecom wavelength of 1550\,nm. Together with the independent work by Ferrari \textit{et al.} \cite{Ferrari_2026}, this work constitutes the first demonstration of polarization-to-time-bin conversion of photons entangled with a single atomic quantum memory. In our work, atom-photon entanglement is generated with a single trapped $^{40}$Ca$^+$ ion and preserved during quantum frequency conversion to the telecom C-band and subsequent polarization-to-time-bin conversion using a fiber-based imbalanced interferometer. The entanglement fidelity after time-bin encoding is evaluated with a second analyzer interferometer and quantum state tomography.

The fidelity of the atom-photon entanglement is measured via full quantum state tomography, which results in $\mathcal{F}_\mathrm{pol}=94.8(0.8)\%$ for the standard polarization encoding ($\mathcal{F}_\mathrm{pol,bg}=94.1(0.8)\%$ without background correction), and $\mathcal{F}_\mathrm{TB}=91.3(4.1)\%$ for time-bin encoding, including the analyzer ($\mathcal{F}_\mathrm{TB,bg}=83.3(4.8)\%$ without background correction). All values include double frequency conversion. The time-bin encoder and analyzer alone have a combined process fidelity of $\mathcal{F}_\mathrm{P}=97.3(1.1)\%$. This is consistent with the factor of 96.3(4.2)\% between the background-corrected entanglement fidelity before and after conversion to time-bin encoding. 

We note that the process fidelity of the time-bin encoder alone will be higher than the measured combined value for encoder and analyzer, provided that it is operated with individual phase stabilization (see \appref{app:phase_stab}).
We also note that the signal-to-background ratio was mainly limited by the high loss in the analyzer interferometer. The raw fidelity (not background-corrected) after the encoder alone will therefore be significantly higher than the measured value for encoder and analyzer. The same increase is expected for the efficiency, or rate, when the encoder is operated as a stand-alone device. 

Through its telecom-wavelength compatibility, the realized system establishes a paradigmatic matter-photon interface for robust long-distance quantum state transmission over optical fibers. At the same time, it provides a key enabling technology for heterogeneous quantum networks, able to interface trapped-ion quantum memories with platforms employing different photonic degrees of freedom. Specifically, it facilitates a heterogeneous quantum repeater architecture combining trapped ions \cite{Bergerhoff_2026} with color centers in diamond \cite{Stolk_2024, Brevoord_2025, Knaut_2024, Schaefer_2025, Herrmann_2026} and demonstrates the feasibility of its deployment on the Saarbrücken urban fiber network \cite{Kucera_2024}. 

\section*{Author contributions}

C.H. and J.G.-F. designed, built and characterized the time-bin conversion setup. 
M.B., C.H. and P.B. prepared the ion experiment.
C.H. performed the measurement. C.H. and J.M. analyzed the data.
T.B. and C.B. provided the QFC technology. 
C.H. and M.B. wrote the paper with input from all authors.
J.E. conceived and supervised the project.

\begin{acknowledgments}
We gratefully acknowledge financial support from the ''Transformationsprogramm Forschung und Wissenstransfer Saar'' through the Center for Quantum Technologies (QuTe) and from the Federal Ministry of Research, Technology and Space (BMFTR) through projects Q.sync (16KISQ045), QR.X (16KISQ001K), QR.N (16KIS2180) and TD.QR (16KISSO21K).
\end{acknowledgments}

\section*{Competing interests}
T.B. and C.B. are involved in developing quantum frequency conversion technology at OPTIQAL Quantum Technologies. The other authors declare no competing interests.

\appendix

\section{Slow polarization drift compensation}\label{app:slow_pol_comp}
Polarization fluctuations in the fibers between the second QFC and the detection setup, as well as between the ion trap and the APC sender stage, are compensated by automatically adjusting the wave-plate angles of the polarization projection setup every $40\,$min. For this purpose, attenuated $854$-nm laser light is sent through the ion trap region and propagated along the single-photon optical path.

Slow polarization drifts in the fibers between the APC receiver and the TBE are compensated using a manual fiber polarization controller, while drifts in the fiber between the TBE and the polarization projection setup are compensated by adjusting the orientation of the projection polarizer. Both manual adjustments are performed approximately once per day.

\section{Temperature stability}\label{app:temp_stab}
\begin{figure}[t]
    \centering
    \includegraphics[width=1\linewidth]{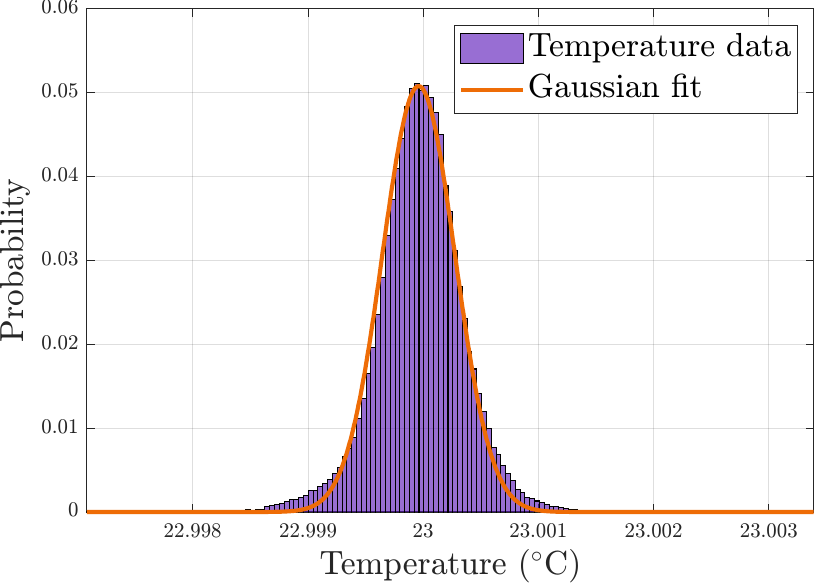}
    \caption{Histogram of the actively stabilized TBE temperature recorded over $55\,$h with a sampling interval of $0.5\,$s together with a Gaussian fit to the measured distribution (orange).}
    \label{fig:temp_hist}
\end{figure}

To quantify the stability of the active TBE temperature stabilization, we individually measure the temperature after reaching thermal equilibrium every $0.5\,$s for around $55\,$h for a set point of $T=23\,^\circ C$, as used in the measurements. The resulting temperature distribution is well described by a Gaussian function, as displayed in \autoref{fig:temp_hist}, with a standard deviation of $\sigma_T=0.44\,$mK.

We estimate the influence of the temperature variation on the transmitted entangled state by first determining the resulting phase fluctuations. The temperature dependence of the travel time $\tau$ in a standard SMF-28 fiber with length $L$ at $\lambda=1550\,$nm was determined in \cite{Slavik2015} as approximately

\begin{align}
    \frac{1}{L}\frac{d\tau}{dT}=39 \,\frac{\text{ps}}{\text{km K}}.
\end{align}

For the encoder path imbalance of $\Delta L=18\,$m this corresponds to a travel time delay $\Delta t$ and resulting temperature-dependent phase sensitivity of

\begin{align}
    \frac{d\Delta \tau}{dT}&= \frac{\Delta L}{L}\frac{d\tau}{dT}=0.70\,\frac{\text{ps}}{\text{K}},\\
\frac{d\phi}{dT}&=\frac{2\pi c}{\lambda} \frac{d\Delta \tau}{dT}=0.85 \,\frac{\mathrm{rad}}{\text{mK}}.\label{equ:phase_temp}
\end{align}

Therefore, applying \autoref{equ:phase_temp} to the measured temperature distribution results in a phase standard deviation of $\sigma_\phi=0.34\,\mathrm{rad}$.

The atom-photon state from \autoref{equ:ape} after transmission through the encoder including the phase variations $\delta\phi$ is written using \autoref{equ:encoder} as

\begin{align}\label{equ:ape_encoder}
\ket{\psi_2(\delta\phi)}=\frac{1}{\sqrt{2}}\left(\ket{a_1}\ket{1}+e^{i(\phi_1+\delta\phi)}\ket{a_2}\ket{2}\right),
\end{align}

with atomic states $\ket{a_1}=1/\sqrt{2}(\ket{+}+\ket{-})$ and $\ket{a_2}=i/\sqrt{2}(\ket{-}-\ket{+})$. For simplicity, the Larmor phase factor of the atomic qubit is omitted in this derivation. The measured ensemble average can then be modeled as

\begin{align}
    \overline{\rho}_2=\int_{-\infty}^\infty P(\delta\phi)\ket{\psi_2(\delta\phi)}\bra{\psi_2(\delta\phi)}\,d\delta\phi,
\end{align}

with the Gaussian phase distribution $P(\delta\phi)$. This yields the quantum state fidelity

\begin{align}
  \mathcal{F}= \bra{\psi_2(0)}\overline{\rho}_2\ket{\psi_2(0)}=\frac{1+e^{-\sigma_\phi^2/2}}{2}.
\end{align}

Assuming no other sources of phase noise or fidelity reduction, such as a transmission imbalance between both interferometer paths, the given phase standard deviation corresponds to a long-term timbe-bin encoder conversion fidelity of $\mathcal{F}=97.2\%$. Since the considered state is maximally entangled, the obtained fidelity is equal to the entanglement fidelity of the photonic channel. Therefore, the estimated reduction is independent of the particular maximally entangled state chosen for this derivation.

\section{Combined encoder and analyzer phase stabilization}\label{app:phase_stab}
The active phase stabilization employs an acousto-optic modulator to generate reference laser pulses of approximately $50\,$ns duration, matching the temporal profile of the single photons. The reference pulses are transmitted through both the TBE and the analyzer interferometer.

Since the stability of the resulting state depends on the relative phase between both setups, as shown in \autoref{equ:analyzer}, we can evaluate the stability of the combined setup via the residual fluctuations of the active stabilization phase signal. The phase variation $\delta\phi$ is inferred from the relative height $h$ of the central peak compared to the side peaks of the resulting three time-bin interference pattern when transmitting the pulsed reference light. Including $\delta\phi$ into the phase acquired by transmission through the long analyzer path, $\phi_2+\delta\phi$, the central peak intensity described by \autoref{equ:analyzer} for a balanced photonic qubit, is given by $I_2=1+\cos(\phi_2-\phi_1+\delta\phi)$. Compared to the side peak intensity, this results in the relative height
\begin{figure}[t]
    \centering
    \includegraphics[width=1\linewidth]{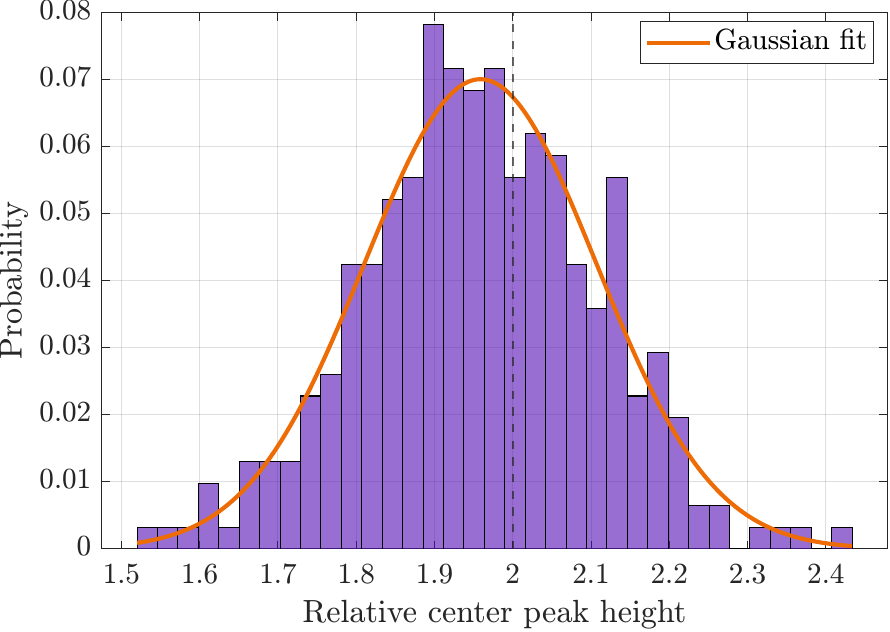}
    \caption{Histogram of the relative height of the central peak compared to the side peaks of three time-bin including a Gaussian fit to the measured distribution (orange) during active phase stabilization to the setpoint $h=2$ (dashed line).}
    \label{fig:center_peak__hist}
\end{figure}

\begin{align}
    h=2\left(1+\cos({\phi_2-\phi_1+\delta\phi})\right).
\end{align}

To ensure maximum sensitivity to phase deviations, we choose $h=2$ as the stabilization locking point, corresponding to the phase difference $\phi_2-\phi_1=\pi/2$. We measure the relative peak height $h$ during active phase stabilization for around $50\,$min. The resulting distribution is displayed together with a Gaussian fit in \autoref{fig:center_peak__hist}, yielding a standard deviation of $\sigma_h=0.15$ and a mean of $\overline{h}=1.96$. Around the locking point, the relative height is described via $h=2\left(1-\sin(\delta\phi)\right)$. Hence, the phase error is given by $\delta\phi=\arcsin(1-h/2)$, resulting in a phase standard deviation of $\sigma_\phi=0.08\,\mathrm{rad}$ and mean locking-point phase offset $\overline{\phi}=0.02\,\mathrm{rad}$.
\begin{figure*}[ht!]
    \centering
    \includegraphics[width=1\linewidth]{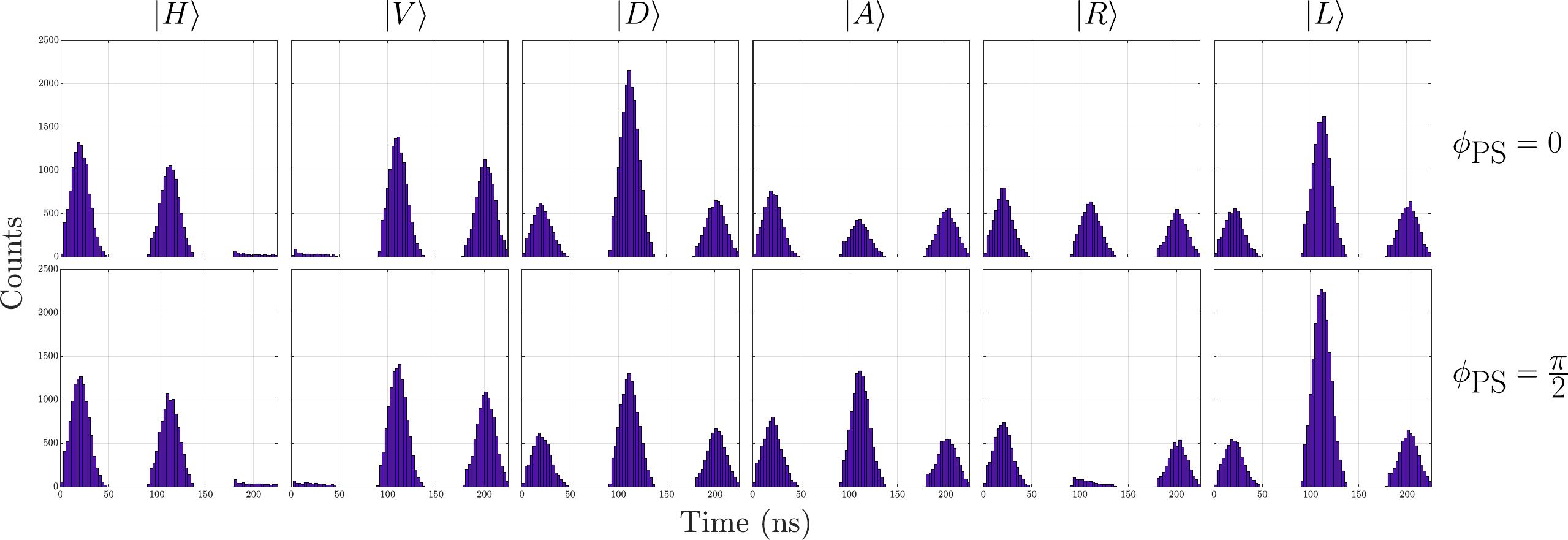}
    \caption{Measurement time-bin signals behind the analyzer for the six inputs polarizations and two phase shifter  phases $\phi_\mathrm{PS}$.}
    \label{fig:TB_Tomo}
\end{figure*}
The influence of the measured phase fluctuations on the atom-photon state is again estimated by \autoref{equ:ape} by including the transmission through the analyzer using \autoref{equ:ape_encoder} and \autoref{equ:analyzer}, resulting in

\begin{align}
\ket{\psi_3(\delta\phi)} = \frac{1}{2}\Bigl[
&\ket{a_1}\ket{1} \nonumber\\
&+\left(e^{i(\phi_2+\delta\phi)}\ket{a_1}
+e^{i\phi_1}\ket{a_2}\right)\ket{2} \nonumber\\
&+e^{i(\phi_1+\phi_2+\delta\phi)}\ket{a_2}\ket{3}
\Bigr].
\end{align}

We model the measured phase error influence as an ensemble average

\begin{align}
    \overline{\rho}_3=\int_{-\infty}^\infty P(\overline{\phi},\delta\phi)\ket{\psi_3(\delta\phi)}\bra{\psi_3(\delta\phi)}\,d\delta\phi,
\end{align}

which for a Gaussian phase distribution $P(\overline{\phi},\delta\phi)$ results in the fidelity

\begin{align}
  \mathcal{F}= \bra{\psi_3(0)}\overline{\rho}_3\ket{\psi_3(0)}=\frac{1+\cos(\overline{\phi})e^{-\sigma_\phi^2/2}}{2}.
\end{align}

Therefore, uncompensated phase fluctuations in TBE and analyzer yield an expected fidelity of $\mathcal{F}=99.85\%$. The remaining discrepancy between this value and the measured process fidelity reported in \appref{app:conv_fid} is primarily attributed to imbalanced transmission probabilities of the beam splitters in the TBE. The higher expected fidelity compared to the estimated fidelity reduction introduced by temperature variations in the TBE demonstrates that the active analyzer stabilization compensates phase fluctuations originating from both the analyzer and the TBE. Consequently, the stand-alone TBE stability, and therefore the achievable time-bin conversion fidelity, can be improved by implementing an active phase stabilization into the encoder.

\section{Losses and background}\label{app:losses} 

We determine the losses of the TBE and analyzer by sequentially adding the individual setups and measuring the transmitted single-photon signal using a $30\,$ns integration window centered on the respective polarization and time-bin-encoded wave-packet centers. The measured average TBE transmission is $45.3\%$. The combined polarizer and analyzer interferometer have a transmission of $8.7\%$. For circularly polarized input light, the latter is ideally limited to $25\%$ by the $50\%$ transmission of the diagonal polarizer and the $50{:}50$ beam splitter in the analyzer interferometer. The additional attenuation from the ideal $25\%$ to the measured $19.2\%$ of the analyzer is attributed primarily to insertion losses into the fiber beam splitters and splice losses between the fiber components.

The measured entanglement rate of $0.460(5)\,$s$^{-1}$ for the polarization qubit atom-photon entanglement is reduced to $0.014(1)\,$s$^{-1}$  when using polarization-to-time-bin conversion and analysis. The observed reduction is largely explained by the measured transmission losses of the inserted optical components.

When connecting the two QFC together with the TBE and analyzer to the detectors, we measure a background rate of $12.3(4)\,$s$^{-1}$. Detector dark counts contribute $2.2(2)\,$s$^{-1}$, hence the remaining, QFC dominated background is $10.1(4)\,$s$^{-1}$.

\section{Conversion fidelity characterization}\label{app:conv_fid}

To characterize the process fidelity of the TBE and analyzer setup, we transmit attenuated laser pulses, generated by an AOM, with a width of $50\,$ns and a delay of $90\,$ns between subsequent pulses. We transmit six different input polarizations $\ket{H}$, $\ket{V}$, $\ket{D}$, $\ket{A}$, $\ket{R}$ and $\ket{L}$ and apply two phase shifter settings $\phi_\mathrm{PS}=0$ and $\phi_\mathrm{PS}=\pi/2$. We record the detector counts for $\approx 25\,$s per input polarization and phase with active phase stabilization every $10\,$s. The respective measured three time-bin signals are shown in \autoref{fig:TB_Tomo}. We measure the counts of all three peaks. The counts $n_0$ and $n_1$ correspond to the early and late time bins, equivalent to the polarization projections $\ket{H}$ and $\ket{V}$. The counts $n_2$ and $n_3$ are obtained from the central time bin for analyzer phases corresponding to projections onto $\ket{D}$ and $\ket{R}$. With $n=(n_0+n_1)/2$ we calculate the Stokes vector components via
\begin{align}
    S_0&=2n,\\
    S_i&=2(n_i-n),\,\,i=1,2,3
\end{align}

and then perform a maximum likelihood estimation reconstruction of the process matrix according to \cite{James_2001}.
From this we calculate the process fidelity $\mathcal{F}_\mathrm{P}=\frac{1}{4}\left(1+\mathrm{Tr}(M)\right)$ from all combinations of three orthogonal Stokes vectors. Averaging all combinations yields the process fidelity $\mathcal{F}_\mathrm{P}=97.3(1.1)\%$.


\bibliography{Bibliography.bib}

\end{document}